\documentclass[aps,prb,preprint,groupedaddress]{revtex4}

\usepackage{graphicx}
\newcommand{\be}{\begin{equation}}
\newcommand{\ee}{\end{equation}}

\begin{document}

\title{Retarded and nonretarded van der Waals interactions 
between a cluster and a 
second cluster or a conducting surface}

\author{M.M. Calbi$^1$, S.M. Gatica$^1$, D. Velegol$^2$ 
and M.W. Cole$^1$}

\affiliation{Departments  of
Physics$^1$ and Chemical Engineering$^2$, Pennsylvania State 
University, University Park, PA 16802}

\date{\today}

\begin{abstract}

In some respects, a cluster consisting of many  atoms may be regarded
as a  single large atom. Knowing  the dielectric properties  of such a
cluster  permits  one  to evaluate  the  form  of  the van  der  Waals
(dispersion) interactions between two  clusters or between one cluster
and a surface. In this paper, we derive these interactions
in  two extreme  opposite regimes  of separation:  fully  retarded and
nonretarded. In the fully retarded regime (very large separation), the
magnitude  of  the  interaction  is  determined  by  just  the  
{\em static}
polarizability  of the  cluster(s). In  the nonretarded  regime (small
separation),  we  employ a  single  resonant  frequency  model of  the
cluster  polarizability to  derive expressions  for  the interactions'
coefficients.  Numerical examples  are presented  to  demonstrate that
many-body  screening of  these  interactions can  be significant.  The
results represent  the corrections to the  commonly used approximation
of pairwise additivity of interatomic interactions.

\end{abstract}

\maketitle

\section{Introduction}

Van  der Waals (vdw) interactions  have been  much studied  in recent
years, in part because of increasing interest in phenomena involving 
surfaces, interfaces, and nanoparticle interactions 
\cite{vdw1,vdw2,book,lucas,israel,met,nano}. In
this paper, we evaluate  two distinct interactions involving clusters,
which  we assume  to be  spherical. One  of these  is  the interaction
between two  clusters, denoted A and  B; the other  is the interaction
between a cluster  and a flat surface. We consider both of these 
interactions in two extreme opposite regimes of separation: 
{\em nonretarded} and {\em fully retarded}, corresponding to small 
separation and very large separation, respectively. The word ``small'' 
is used advisedly because we assume throughout this paper that the 
separation is large compared to the size of the clusters.

Nonretarded vdw interactions, sometimes called {\em London forces}, 
can be derived from quantum mechanics and electrodynamics. Most 
research involving interactions within condensed states of matter 
pertains to the nonretarded regime because short-range forces 
predominate in determining the physical behavior of such systems. 
However, the onset of effects of retardation occurs at sufficiently 
small distance scales ($\approx$ 10 nm) that such effects are 
often present to some extent; they become particularly important 
when studying large molecules or clusters, because of their size. 
The interaction is said to be {\em fully retarded}$\,$\cite{marg,cp,dlp} 
in the separation regime where the finite speed of light causes a 
significant attenuation of the (nonretarded) vdw interaction.
Indeed, even the power law
describing the distance dependence of the interaction is altered by retardation (increasing the falloff by one power of distance). The regime  of
retardation is that where the  separation $r$ is large compared to both
the cluster size  (radius $R$) and a {\em typical}  wavelength $\lambda$ 
determined by
the  spectra  of both  interacting  species: $\lambda=hc/\Delta E$;  
here $\Delta E$  is  a
characteristic energy in the electronic spectrum of the cluster and/or
surface. If $\Delta E$ is the first ionization energy, for  example, 
then $\lambda \approx$ 0.1 to 1 micron (for Na and Si, 
$\lambda \approx$ 0.24 microns and 0.15 microns, respectively). 
The surface
force apparatus has been used to measure van der Waals
forces between two macroscopic mica surfaces \cite{mica}. Other 
techniques (e.g., atomic force microscopy \cite{atm} and total 
internal reflection microscopy \cite{int} have been used to measure 
interaction forces, and a few techniques (e.g., laser trapping 
\cite{laser} and differential electrophoresis \cite{darrell}) have 
been used to measure sub-piconewton interaction forces between
Brownian particles. Figure 1 presents results for the 
interaction between two Na clusters obtained 
in this paper for the nonretarded and fully retarded regimes, along with 
a plausible interpolation over the interval $0.1 < r/\lambda < 10$.

Because vdw interactions arise from fluctuating electromagnetic fields
in  the interacting  media, there  arise many-body  effects associated
with screening of these interactions by neighboring particles within a
cluster.  This   means  that  one  should not  simply   add  together  
the
interactions of individual particles without taking the screening into
account \cite{ham}. As an example, we  demonstrate the importance of screening in
the case of alkali metal  clusters. There occurs a general increase in
the effect of  screening as a function of cluster  size, but the trend
is not completely monotonic because of electronic shell effects on the
cluster's polarizability.

The outline  of this paper is  the following. Section  II describes the
case of fully retarded interactions. In that case, the vdw interaction
is  determined by just  the static  polarizability of  the interacting
clusters   (because  long   wavelength   fields  give   rise  to   the
interaction).  In Section III,  we explore  the nonretarded  regime, for
which  all  wavelengths   contribute,  in  principle.  That  treatment
requires the introduction of a  model form for the polarizability as a
function of  frequency. In  the present case,  we employ  the simplest
possible  model (single  resonant frequency)  that is  consistent with
both  the static  value  and the  known  high frequency  limit of  the
polarizability. Section IV summarizes our results and comments about possible extensions to related problems.

\section{Fully retarded regime}

The starting point of our analysis is the fact that the fully retarded
interaction  between two  atoms (here  denoted $a$ and $b$)  separated by
distance $r$ (with $r \gg \lambda$) is known to satisfy \cite{marg}:

\be
V_{ab}(r) = - K \frac{P_a^0 P_b^0}{r^7} \;\; , \;\; K=\frac{23 hc}{8 \pi^2}
\ee

Note the interesting fact that the interaction falls off with distance
 as  $r^{-7}$,   with  a  coefficient  determined  by   the  
{\em static}  atomic
 polarizabilities  ($P^0_a$, $P^0_b$)  of  the  species  involved.  
This  behavior
 contrasts with  the nonretarded vdw interaction in  both the latter's
 power  law  ($r^{-6}$)  and  its  dependence  on  the  
frequency-dependent polarizability, rather than just the static value 
$P^0$ (see Eqs. \ref{vnr} and \ref{c6atom} below). 

\subsection{Cluster-cluster interaction}

We  now   turn  to  the   problem  of  the  interaction   between  two
clusters.  The key  idea of  this  paper is  simple: one  may treat  a
cluster as  though it is a large  atom, insofar as the  cluster has an
excitation  spectrum  with   an  energy  gap  and  it   is  of  finite
extent. Here, we consider the case of separation $r$ large compared to its
size.  Specifically,   the  present  calculation   requires  that  the
cluster's radius $R \ll \lambda \ll r$. Validity of that 
criterion permits  
one to use
the  local,  dipolar  approximation  to  describe  the  electrodynamic
response of the cluster to the fluctuating electromagnetic fields that
are  responsible for  vdw interactions.  Based on  this  argument, the
interaction between two clusters A and B may be written

\be
V_{AB}(r) = - K \frac{P_A^0 P_B^0}{r^7} 
\ee

Here $P_A^0$ and $P_B^0$  are the static polarizability of cluster $A$ 
and $B$, respectively, a subject
of much  attention in recent years \cite{rmp,Na,knight,kresin}. 
Now, one  may ask a
simple  question: how  different is  this  cluster-cluster interaction
from  a   naive  estimate, $V_{est}$, obtained  from   summing  pairs  of
interactions between  the constituent a, b atoms? The latter  may be
written  as a  product of  the interatomic pair interaction  and the
number of such interactions between the two clusters:

\be
V_{est}(r) = N_A N_B V_{ab}(r) = - K N_A N_B \frac{P_a^0 P_b^0}{r^7}
\ee

We compute  the ratio  of the  exact result Eq. (2) to this  estimate and
define the ratio as a ``screening'' function $S_{AB}^r$:

\be
S_{AB}^r(r) = \frac{V_{AB}(r)}{V_{est}(r)} = \frac{P_A^0(N_A)}{N_A P_a^0}\,
\frac{P_B^0(N_B)}{N_B P_b^0} = f_A(N_A) f_B(N_B)
\label{scret}
\ee

This  screening function is separable, a product  of  two
independent screening  functions $f(N)$, each  of which is the  ratio of
the  cluster's  polarizability to  a  nominal cluster  polarizability,
equal to  the product of the  atomic polarizability and  the number of
atoms. Because  each of these  $f$ functions is  less than one  
if $N > 1$, then $S_{AB}^r < 1$  if  either  of  the clusters  
possesses  more  than  one atom.  Thus the  true interaction  is  reduced 
relative  to the  naive estimate $V_{est}$. 

Figure 2 presents the screening function  for the case of a Na cluster
interacting with a second Na cluster. The static polarizalility of the
Na clusters $P^0 (N)$ was taken from theoretical
results, obtained using the jellium model and density-functional 
calculations \cite{Na}. This simple model was adequate to explain 
the experimental polarizability data \cite{knight}. 
For the  extreme case $N_A = 1 = N_B$, the screening  function is
unity, by definition. For  increasing N, typically, the screening causes
$S_{AB}^r$ to decrease below one. 
Note that the reduction  is by a 
factor near 0.4 for two 40 atom Na clusters and this is 
essentially the asympotic value \cite{asy} for $N_A$, $N_B \rightarrow \infty$. 
The dependence on number is not monotonic,
however, because  of shell structure  of the electronic states  of the
cluster.  This  nonmonotonic behavior  makes  the figure  interesting;
notice the shape of a ``bat head'' in the center of the Figure 2.

We remark that  the hypothetical case of clusters  comprised of weakly
polarizable,  inert  gases  represents   an  instance  for  which  the
functions  whose product  determines $S_{AB}^r$  are  close to  1, 
i.e.  $P_A^0=N_A P_a^0$. Indeed, in the  framework of local, 
continuum electrostatics, this
relation  is exact  if  the Clausius-Mossotti  relation describes  the
static dielectric constant $\epsilon$ of the material, since the 
dipole moment of
a dielectric  cluster \cite{jac} in  an external field {\bf E} is  
$[(\epsilon-1)/(\epsilon+2)] R^3 \bf{E}$.
Thus,  the interaction  between two  such  inert gas  clusters is 
{\em not}
significantly screened.

\subsection{Cluster-surface interaction}

Next,  we address  very briefly  the case  of a  cluster (A)  near the
surface of  a semi-infinite perfectly  conducting metal, a  simplified version  of a
more  general adsorption  problem.  In  that case,  the
cluster's interaction with the surface obeys a relation \cite{book}

\be
V_{cond} = - K' \frac{ P_A^0}{z^4}
\label{cond}
\ee

Here $K'= 3 hc /(16 \pi^2)$ and  $z$ is the  separation between  
the cluster's center and the boundary of the surface . The  ratio of 
this  cluster-surface interaction to a  naive estimate,
based on $N_A$ individual constituent atoms of the  cluster is analogous
to the ratio $S_{AB}^r$ for the cluster-cluster interaction:

\be
\frac{V_{cond}}{V_{est}}= \frac{P_A^0(N_A)}{N_A P_a^0}=f_A(N_A)
\ee

\noindent This  behavior therefore involves  the same  function as  the previous
problem \cite{klim}.

Finally, we  evaluate the retarded cluster-surface  interaction in the
case of  a solid made of  molecules of static  polarizability $P_m^0$ 
and density $n$.  We specialize  the discussion to  the situation  
where the
product $n P_m^0 \ll 1$; this weak-screening limit is essentially the 
extreme
opposite case from the perfect conductor considered above. In this new
limit,  the interaction  $V_{weak}$  can be  evaluated  by 
integrating  the
cluster-molecule interaction density $- K n  P_A^0 P_m^0/r^7$ over the 
half-space $z<0$ in the  presence of the cluster, centered at 
distance $d$ above the
surface:

\be
V_{weak}(d)=-n K P_A^0 P_m^0 \int_{z'< 0} \frac{d{\bf r'}}{|\hat{\bf z} d 
- {\bf r'}|^7} 
\ee

The result of that integration is:

\be
V_{weak}(d)= - \frac{\pi}{10} K n \frac{P_A^0 P_m^0}{d^4} = - \frac{23}{80 \pi}\frac{h\, c\, n\, P_m^0 P_A^0}{d^4}
\ee

We may  compute the ratio of  this interaction to that  of the perfect
conductor with the same cluster (Eq. (\ref{cond})):

\be
\frac{V_{weak}}{V_{cond}} = \frac{23}{15}\, \pi n P_m^0
\label{weak}
\ee

Note that these interactions strengths  are similar if 
$n P_m^0 \approx 0.2$. Even
for  such  a  small  polarizability,  therefore,  the  interaction  is
comparable to that obtained with a perfect conductor. In the case of an 
Ar solid, in contrast, the ratio in Eq. (\ref{weak}) is about 0.2.

\section{Nonretarded interaction}

We first recall that the 
nonretarded interaction between two atoms $a$ and $b$ separated by
 a distance $r$ is given by \cite{marg}

\be
V_{ab}(r)= - \frac{C_{6}^{ab}}{r^6}
\label{vnr}
\ee

\noindent where the constant $C_{6}^{ab}$ is 

\be
C_{6}^{ab}=\frac{3 \hbar}{\pi} \int_0^{\infty} du P_a(iu)\,
P_b(iu)
\label{c6atom}
\ee

Here, $P_a(\omega)$ and $P_b(\omega)$ are the dynamic polarizabilities 
of the atoms, continued to imaginary frequencies $\omega=iu$.
The atomic polarizability has the form

\be
P_{a}(iu)= \frac{e^2}{m} \sum_n \frac{f_{0n}}
{\omega_{0n}^2+u^2}
\ee

\noindent where $f_{0n}$ is the oscillator strength that measures the probability of the transition from the ground state $0$ to the excited state 
$n$, at frequency $\omega_{0n}$. 

\subsection{Cluster-cluster interaction}

To compute the interaction between two clusters A and B, we use the same
idea as in the last section of considering the clusters as large atoms 
($R \ll r \ll \lambda$). In that case, we may write the cluster-cluster interaction as:

\be
V_{AB}^{nr}(r)= - \frac{C_6^{AB}}{r^6} \;\;\; , \;\; 
C_6^{AB} = \frac{3 \hbar}{\pi} \int_0^{\infty} du P_A(iu)\,
P_B(iu) 
\ee
 
We propose a simple form for the dynamical polarizability 
of the cluster \cite{drude}: 

\be
P_A(iu)=\frac{P_A^0}{1+\frac{u^2}{\omega_A^2}}
\label{dynp}
\ee  

\noindent designed to give the static polarizability of the cluster 
at zero frequency, and the known asympotic behavior, 
$P_A \rightarrow Z N_A e^2/(m u^2)$ in the high frequency limit (free electrons) if we define $\omega_A$ with

\be
\omega_A^2 \equiv \frac{Z N_A e^2}{m P_A^0}
\ee

Here $m$ is the electron mass and $Z$ is the number of electrons in the 
atom. An analogous aproximation can be made for the atomic polarizability 
$P_a(iu)$ using a single characteristic frequency $\omega_a$ to determine its form. The characteristic frequency of the cluster, $\omega_A$, is 
larger than the 
corresponding result for a single atom, $\omega_a$, by a factor of 
$1/\sqrt{f_A}$. We have tested the quality of our proposed function
by comparing it with the dynamical 
polarizability of small metal clusters, calculated by Kresin 
within the random-phase approximation \cite{kresin}. He finds 
two collective excitations 
(one surface mode and one volume mode) in terms of which he evaluates 
the dynamical polarizability. We 
have found that the overall frequency dependences of his expression and 
Eq. (\ref{dynp}) agree very well, with a difference less than 0.5 $\%$.

Assuming the cluster polarizability to be that from 
Eq. (\ref{dynp}) (and using the corresponding approximation for the 
atomic polarizability), we compute the ratio $S_{AB}^{nr}$ between the 
nonretarded cluster-cluster interaction and the simple estimate 
that results 
from summing pair interactions, assuming $Z=1$:

\be
S_{AB}^{nr}=\frac{C_6^{AB}}{N_A N_B C_6^{ab}}= \frac{\omega_a \omega_b 
(\omega_a + \omega_b)}{\omega_A \omega_B 
(\omega_A + \omega_B)}=\sqrt{f_A \, f_B} \; 
\frac{(P_a^0)^{-1/2}+(P_b^0)^{-1/2}}{(f_A P_a^0)^{-1/2}+
(f_B P_b^0)^{-1/2}}
\ee

Figure 3 shows this function for the case of two interacting Na
 clusters. It shares the same qualitative behavior as the screening 
function in the retarded case (Eq. (\ref{scret})). In the case $a=b$, 
$N_A=N_B$, 
$S_{AB}^{nr}=f_A^{3/2}(N_A)$. This compares with the result $f_A^{2}$ in the fully retarded limit (Eq. (\ref{scret})). 

\subsection{Cluster-surface interaction}
 
The nonretarded vdw interaction between an atom and a surface at separation $d$ is given by \cite{book}

\be
V_{a-surf}(r)= - \frac{C_{3}^{a-surf}}{d^3}
\ee

\noindent with

\be
C_{3}^{a-surf}=\frac{\hbar}{4 \pi} \int_0^{\infty} du P_a(iu) g(iu)
\ee

Here, $g(iu)$ is the dielectric response function of the substrate 

\be
g(iu)=\frac{\epsilon(iu)-1}{\epsilon(iu)+1}
\ee

\noindent which can be approximated by

\be
g(iu)=\frac{g_0}{1+\frac{u^2}{\omega_s^2}}
\label{ans}
\ee

\noindent if the response is dominated mainly by the resonance 
at $\omega_s$ (the surface plasmon resonance in the case of a metal).
 The parameter $g_0$ is equal to 1 in the case of a free electron metal,
in which case the ansatz Eq. (\ref{ans}) is exact, with $\omega_s = \sqrt{2}$ 
times the bulk plasma frequency. 
Following the spirit of the previous sections, the van der Waals 
coefficient for the cluster-surface problem is 

\be
\frac{C_{3}^{A-surf}}{N_A}=\frac{\hbar}{4 \pi} \int_0^{\infty} du \frac{P_A(iu)}{N_A}\, g(iu)=\frac{g_0 f_A P_a^0}{8} \frac{\hbar \omega_s}
{\left( 1+\frac{\omega_s}{\omega_a}\right)} 
\ee

After carrying out the integrations for $C_{3}^{A-surf}$ and 
$C_{3}^{a-surf}$, the
screening function for the cluster-surface interaction results:

\be
S_{A-surf}=\frac{C_3^{A-surf}}{N_A C_3^{a-surf}}=f_A\; 
\frac{1+\frac{\omega_s}{\omega_A}}{1+\frac{\omega_s}
{\omega_A}\sqrt{f_A}}
\ee 

We plot this function in Figure 4, for different ratios 
$\omega_s/\omega_A$. As can be seen in the figure, $S_{A-surf} 
\rightarrow \sqrt{f_A}$, if $\omega_s \gg \omega_A$, and goes to
$f_A$ when  $\omega_s \ll \omega_A$. This relative interaction ratio 
is the smallest in the latter case (since $f_A < 1$).

\section{Summary}

In this paper we have evaluated two kinds of interaction 
(cluster-cluster and cluster-surface) in two extreme opposite regimes of 
separation (nonretarded and fully retarded).

We  have described a particularly simple result for the  fully
retarded  interactions  involving  clusters. The  interaction  between
two clusters is reduced  (relative to the naive estimates) 
by a screening
function, derived from  the product  of individual  cluster screening
functions  $f(N)$, that  depend  on the  electronic  properties of  the
cluster. Because  the latter shows  interesting, nonmonotonic behavior
as  a function  of atomic  number,  the dependence  on the  individual
clusters' numbers is  nontrivial,  as exemplified  in  Figure 2. 
Qualitatively similar behavior occurs in the nonretarded case, seen 
in Figure 3.

This work can be generalized in many ways. One potential extension 
involves exploration of the intermediate regime of separation. While 
straightforward in principle, this is complicated in practice, so we 
simply mention results obtained elsewhere for analogous problems in 
this intermediate separation regime (the dashed region in Figure 1). 
In the case of the interaction 
between two identical atoms, the effect of retardation is to reduce 
the nonretarded interaction by a factor of two when the atoms are 
separated by one-tenth of the characteristic wavelength in their 
excitation spectrum (i.e., r is about 40 nm if the relevant energy 
is 3 eV \cite{ref}). In  the case of the interaction 
between an inert gas atom, or small molecule, and a surface, the 
corresponding distance is somewhat smaller, about 20 nm \cite{ads}. 
These distances are sufficiently small that one should not ignore the 
effects of retardation in many applications. If one were to accept the 
linear interpolation in Figure 1, the ratio of this interaction to the 
(extrapolated) nonretarded interaction at $r=\lambda$ would be 0.49.

The second problem of potential interest is the effect of a medium 
in which the clusters are dissolved. This is the principal focus 
of our future research. A third problem of potential interest is 
larger clusters (i.e. radii {\em not} much smaller than their 
separation). Both of these present no question of principle but 
neither problem is as simple computationally as the problems 
described here.

\begin{acknowledgments}

We are grateful to A.W. Castleman, K.A. Fichthorn, G. Holtzer, and 
A.A. Lucas for helpful communications 
concerning this work and would like to thank the National Science 
Foundation and the Ben Franklin Technology Center for their support.

\end{acknowledgments}

\newpage

\begin{figure*}
\includegraphics[height=4in]{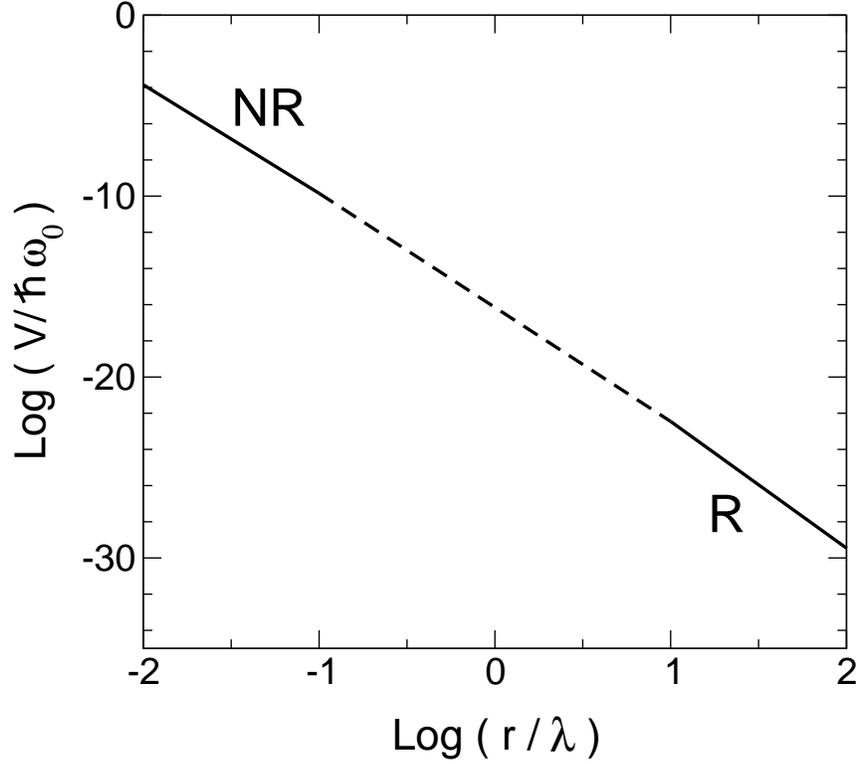}
\caption{Different regimes of the van der Waals interaction between
two 20 Na atoms clusters, in units of a characteristic energy 
$\Delta E=\hbar \omega_0$. The nonretarded regime (NR) when 
$R/\lambda \ll r/\lambda \ll 1$ is analyzed in Section III, and the 
retarded regime for separations such that 
$r/\lambda \gg 1 \gg R/\lambda$ is explored in Section II. The 
intermediate regime (dashed line) is not considered in this work,
but note that it can be interpolated between the $r^{-6}$ 
(nonretarded) and $r^{-7}$ (retarded) functional forms.}
\end{figure*}

\begin{figure*}
\includegraphics[height=7in]{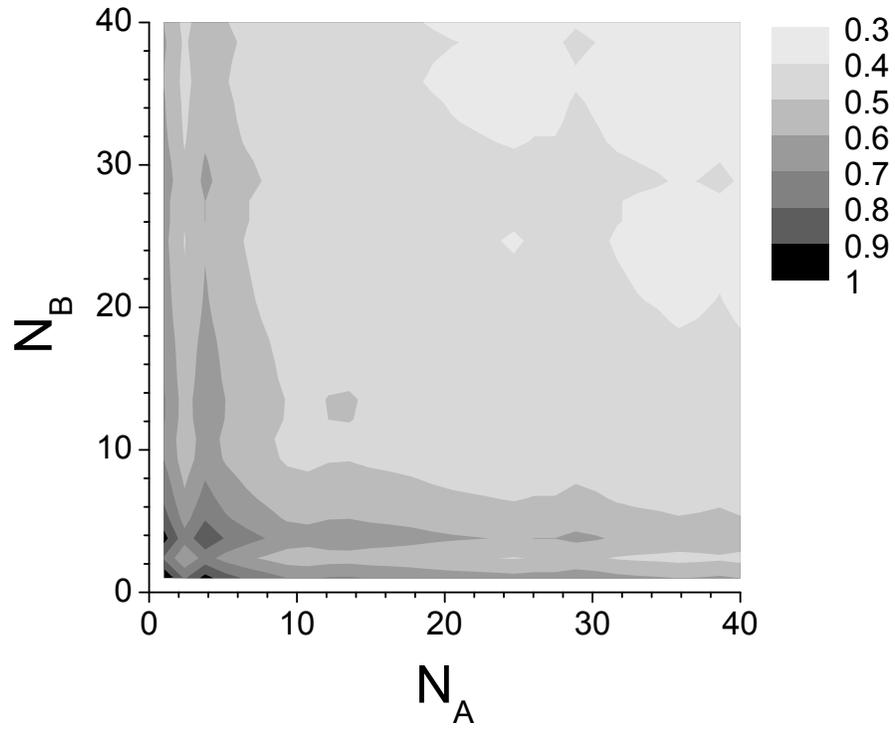}
\caption{Retarded screening function $S^r(N_A,N_B)$ for the case 
of two Na clusters containing $N_A$ and $N_B$ atoms, respectively.}
\end{figure*}

\begin{figure*}
\includegraphics[height=7in]{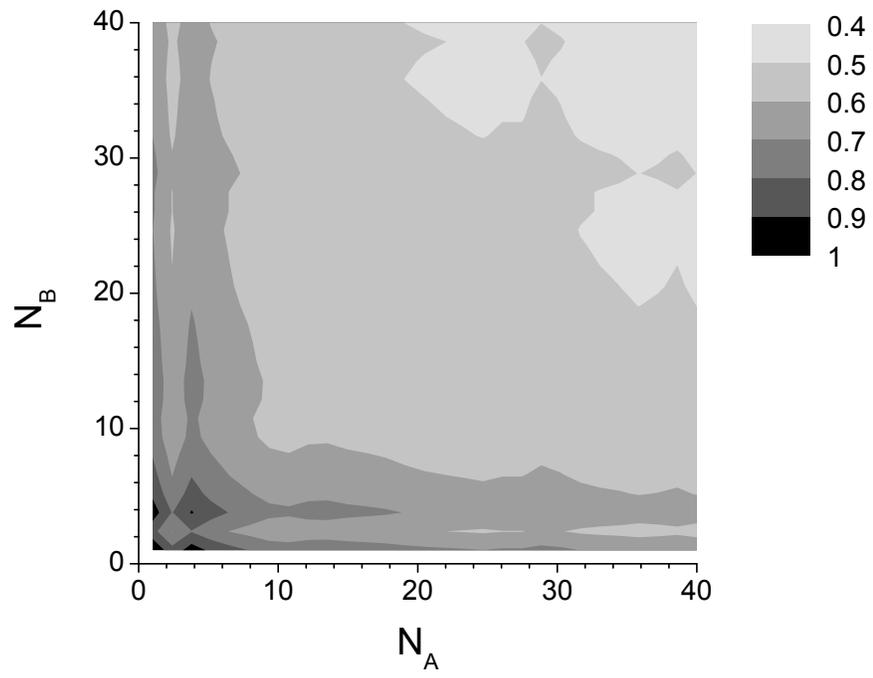}
\caption{Same as Fig. 1 for the nonretarded interaction.}
\end{figure*}

\begin{figure*}
\includegraphics[height=4in]{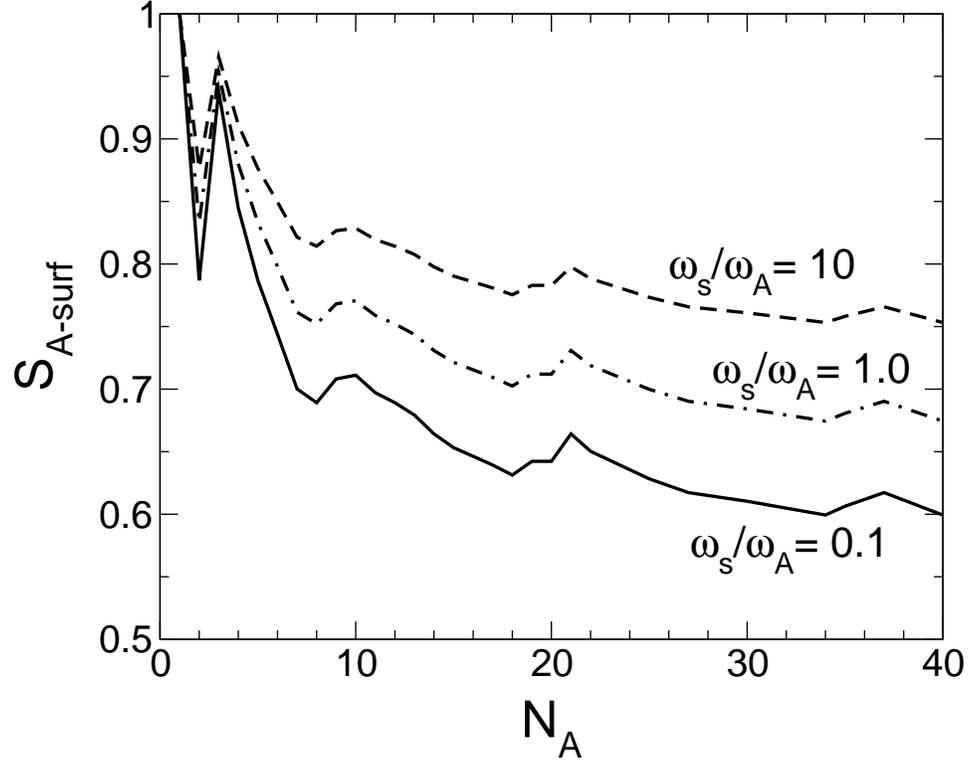}
\caption{Nonretarded screening function $S_{A-surf}$ for the case 
of a Na cluster 
of $N_A$ atoms interacting with a surface which has a characteristic
 energy $\hbar \omega_s$.}
\end{figure*}


\begin{references}
\bibitem{asy} This asymptotic value is a consequence of the behavior 
of the individual screening functions $f(N)$ for large $N$, that 
decrease from unity for N=1 until they reach an aproximately constant 
value $f(N) \approx 0.6$ for $20<N<40$.
\bibitem{klim} We note that G.L. Klimchitskaya, U. Mohideen and 
V.M. Mostepanenko (Phys. Rev. A {\bf 61}, 062107, 2000) have derived 
a different 
expression for a related quantity. Their result  is determined for 
the case when $R$ greatly exceeds the  distance to the surface, which 
itself
is diverging. The  present case involves finite $R$,  which is assumed 
to be much smaller than the distance $z$.
\bibitem{drude} Such a single frequency approximation has been widely 
used in computing interatomic interactions and adsorption interactions 
(see Refs. 3, 4 and 5).
\end{references}
\end{document}